\documentclass[twocolumn, 10pt]{article}
\usepackage{amsmath,amssymb}
\usepackage{graphicx}
\usepackage[format=hang,labelfont=bf,textfont=small,singlelinecheck=false,justification=raggedright,margin={12pt,12pt},figurename=Fig.]{caption}
\usepackage{titlesec}
\usepackage{textcomp}
\usepackage{apacite}
\usepackage[authoryear]{natbib}
\usepackage{hyperref}
\usepackage{booktabs}

\makeatletter
\def\affiliation#1{\gdef\@affiliation{#1}}
\def\abstract#1{\gdef\@abstract{#1}}
\def\graphabst#1{\gdef\@graphabst{#1}}
\def\keywords#1{\gdef\@keywords{#1}}
\def\corresp#1{\gdef\@corresp{#1}}

\newcommand{\MakeTitle}{
  \newpage
  \null
  \vskip 2em%
  \begin{center}%
  \Large \@title\par
  \vskip 1em%
  \large \@author
  \end{center}
  \noindent\@affiliation\par
  \vskip 1em%
  \noindent\@corresp\par
  \vskip 1em%
  \noindent\@abstract\par
  \noindent\@graphabst\par
  \vskip 1em%
  \noindent\@keywords\par
}
\makeatother

\newcommand*{\TitleFont}{%
      \usefont{\encodingdefault}{\rmdefault}{}{n}%
      \fontsize{18}{12}%
      \selectfont}

\titleformat{\section}
  {\normalfont\fontsize{10}{11}\bfseries}{\thesection.}{2pt}{}
  \titlespacing*{\section}{0pt}{12pt}{6pt}

\titleformat{\subsection}
  {\normalfont\fontsize{10}{10}\bfseries}{\thesubsection.}{2pt}{}
  \titlespacing*{\subsection}{0pt}{6pt}{0pt}
  
\titleformat{\subsubsection}
  {\normalfont\fontsize{10}{10}\bfseries}{\thesubsubsection.}{2pt}{}
  \titlespacing*{\subsubsection}{0pt}{6pt}{0pt}
  
\normalsize

\title{\TitleFont Data Science in Biomedicine}
\author{Yovaninna Alarc\'on-Soto$^{1\ast}$, Jenifer Espasand\'in-Dom\'inguez$^{2}$, Ipek Guler$^{3}$, Mercedes Conde-Amboage$^{4}$, Francisco Gude-Sampedro$^{5}$, Klaus Langohr$^{1}$, Carmen Cadarso-Su\'arez$^{2}$, Guadalupe G\'omez-Melis$^{1}$}
\affiliation{$^{1}$Departament d'Estad\'istica i Investigaci\'o Operativa, Universitat Polit\`{e}cnica de Catalunya/ BARCELONATECH, C/ Jordi Girona, 1--3, 08034 Barcelona, Spain\\
$^{2}$Unit of Biostatistics,	Department of Statistics, Mathematical Analysis, and Optimization, Universidade de Santiago de Compostela\\
$^{3}$Instituto Maim\'onides de Investigaci\'on Biom\'edica de C\'ordoba (IMIBIC)\\
$^{4}$Models of Optimization, Decision, Statistics and Applications Reseach Group (MODESTYA), Department of Statistics, Mathematical Analysis, and Optimization, Universidade de Santiago de Compostela\\
$^{5}$Clinical Epidemiology Unit, Complejo Hospitalario Universitario de Santiago de Compostela\\}
\corresp{$^{\ast }$Corresponding author yovaninna.alarcon@upc.edu, Tel.:$+$34 934 016 947; fax: $+$34 934 015 855}

\abstract{\textbf{Abstract}: We highlight the role of Data Science in Biomedicine. Our manuscript goes from the general to the particular, presenting a global definition of Data Science and showing the trend for this discipline together with the terms of cloud computing and big data. In addition, since Data Science is mostly related to areas like economy or business, we describe its importance in biomedicine. Biomedical Data Science (BDS) presents the challenge of dealing with data coming from a range of biological and medical research, focusing on methodologies to advance the biomedical science discoveries, in an interdisciplinary context.}

\graphabst{
\begin{center}
\end{center}
}

\begin{document}

\onecolumn
\MakeTitle

\section{Introduction}
\label{sec:intro}
In the last 10 years, we have observed an important increase in the number of job offers requesting data scientists. Data science  was already recognized as a science more than 5 decades ago by John Tukey. In the article \textit{The Future of Data Analysis} he points out that more emphasis should  be placed on using data to suggest hypotheses to test and reflects on the existence of an as-yet unrecognized science, whose subject of interest was learning from data \citep{donoho201750} and that lays the foundation of today \textit{data science} area. \lq Data analysis', includes
\begin{quotation}
\lq\lq (\ldots) among other things: procedures for analyzing data, techniques for interpreting the results of such procedures, ways of planning the gathering of data to make its analysis easier, more precise or more accurate, and all the machinery and results of (mathematical) statistics which apply to analyzing data.'' \citep{tukey1962future}
\end{quotation}

Due to the technological explosion of the last few years, massive amounts of data are generated every day in different areas. This new era requires the development of new techniques to analyze and draw reliable conclusions from these data. In this context, the figure of the data scientist emerges, proclaimed by \citet{davenport2012data} as \lq the Sexiest Job of the 21st Century'. But, what exactly is a data scientist?

This question has been already addressed by many other researchers, such as \citet{schutt2013doing} or \citet{donoho201750}, and it has been the topic of many columns and discussions in important media such as The Guardian or The New York Times.

To provide a definition of data science in our own terms, we start by refering to the definition of data scientist found in the Oxford Dictionary \citep{dictionary2008oxford}:
\begin{quotation}
\lq\lq A person employed to analyze and interpret complex digital data, such as the usage statistics of a website, especially in order to assist a business in its decision-making.''
\end{quotation}

\begin{figure}[!h]
\centering
\includegraphics[scale = 0.4]{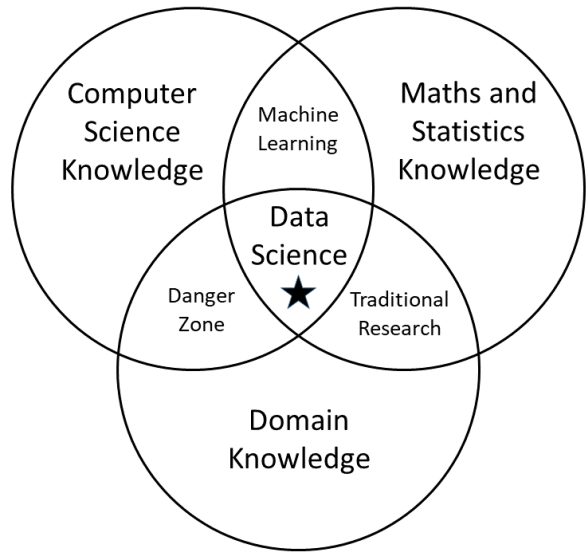}
\caption{Data science scheme based on the Conway's Venn diagram \citep{conway2010data}.}
\label{fig:conwaysvenn}
\end{figure}

We will follow the very helpful data science scheme created by \citet{conway2010data} to  explore the different attributes a data scientist should convey (Figure \ref{fig:conwaysvenn}).  First, knowledge in Mathematics and Statistics is necessary.  Mathematics gives  a universal language and is essential for solving real-world problems. From Statistics comes the understanding and experience to work with data, selecting the appropriate techniques to deal with it, to pre-process, summarize, analyze and draw conclusions. Second, computer science know\-ledge is also fundamental.  Not only getting computers to do what you want them to do requires intensive hands-on experience, but also computer scientists must be adept at modelling and analyzing problems. They must also be able to design solutions and verify that they are correct. Problem solving requires precision, creativity, and careful reasoning. Computer science has a wide range of sub-areas. These include computer architecture, software systems, graphics, artificial intelligence, computational science, and software engineering. Drawing from a common core of computer science knowledge, each of these areas focuses on particular challenges.

The third and not least important characteristic of a data scientist is domain knowledge, a thorough understanding of the field in which the research is being developed is needed to understand the research context and more important to be able to provide realistic and responsible answers to the questions at hand. Examining what these three areas have in common, the intersection between mathematical and statistical knowledge and domain knowledge is the most common, from which traditional research emerges, whereas Machine Learning arises from the intersection of mathematical and statistical knowledge and computer science knowledge. The name Machine Learning, coined by \citet{samuel1959some}, is a field of computer science that uses statistical techniques to give computer systems the ability to learn with data. Nevertheless, if there is not enough statistical knowledge to choose the appropriate methods and analyses for the pertinent research objectives, mixing expertize in the field of research with computer science knowledge might lead us to a danger zone. 

This overlap of skills gives people the ability to create what appears to be a legitimate analysis without any understanding of how they got there or what they have created \citep{conway2010data}. As \citet{wilson1927statistics} stated
\begin{quotation}
\lq\lq (\ldots) it is largely because of lack of knowledge of what statistics is that the person untrained in it trusts himself with a tool quite as dangerous as any he may pick out from the whole armamentarium of scientific methodology.''
\end{quotation}

We believe, however, that further soft skills are required by a data scientist. For this reason, we have added a star in the intersection of the three areas, in the core of the Data Science concept.  A data scientist needs not only to be an expert in his or her area, but also a good communicator, collaborator, leader, advocate, and scholar. As a communicator, the key competencies are active listening and asking questions, explaining advantages or shortcoming of statistical and computer methods, and interpreting results in a meaningful way in the context of the application. He or she has to be a fine collaborator, because he or she will have to work in  interdisciplinary teams. In addition, being a  leader is the key to successfully influence multidisciplinary research, the data scientist will have to advocate to use his or her expertise, and given that science is continuously developing, a data scientist has to be a scholar. In a recent paper by \citet{zapf2018makes} these soft skills are already identified for being a successful biostatistician, and they can be generalized to any data scientist.

Therefore a data scientist needs to master a set of skills ---mathematical, statistical, computational, communication skills--- that are not easy to develop for a single person. Given the scarcity of people with such a complete profile, there is a need to create multidisciplinary working groups formed by different specialists who add their qualities to make room for data science itself.

The article is organized as follows: in Section~\ref{sec:dsglobalimpact}, we analyze the global impact of Data Science by updating the research of \citet{kane2014cleveland} in which the author analyzes the search-term usage of \lq\lq Data Science'' over time until 2014 adding \lq\lq Cloud Computing'' and \lq\lq Big Data'' to the search, until 2018 and using Google Trends. This section includes an overview of the Data Science journals. Following, in Section~\ref{sec:biomedical}, we describe Data Science in Biomedicine, or Biomedical Data Science (BDS), present a web search restricted to the biomedical area, and include some examples of BDS studies. Finally, the main findings of this work are summarized in Section~\ref{sec:discus}. 

\section{Data Science: Global Impact and dissemination}
\label{sec:dsglobalimpact}

\citet{cleveland2014data} proposes an action plan for statistics, in which he elevates the role of the statistician to the level of a researcher who should not limit him or herself to providing only statistical calculations and p-values, but should, also, be involved in the interpretation of these.

Data science has become very popular in recent years as a tool in many fields such as Economics (business analytics, fraud and risk detection), internet search, digital advertisements, image and speech recognition, delivery logistics, gaming, price comparison websites, airline route planning, robotics, among others. To contextualize the impact of this new discipline all over the world, we have used Google Trends to update the research of \citet{kane2014cleveland}. Kane analyses the search-term usage of \lq\lq Data Science'', \lq\lq Cloud Computing'' and \lq\lq Big Data'' until 2014 (see Figure \ref{fig:conwaysvenn}). \lq\lq Cloud Computing'' and \lq\lq Big Data'' were added because of their close relation with Data Science, their intrinsic relation with the computational techniques and to frame the evolution of the impact of the Data Science. It must be taken into account that Google Trends is an online search tool that allows the user to see how often specific keywords, subjects, and phrases have been queried over a specific period of time and provides information about Google searches all over the world. Search trends show how the interest for a given term has evolved over time by assigning a score between 0 and 100 to search terms on a year-by-year basis. 

\begin{figure}[!ht]
\centering
\includegraphics[width = \textwidth]{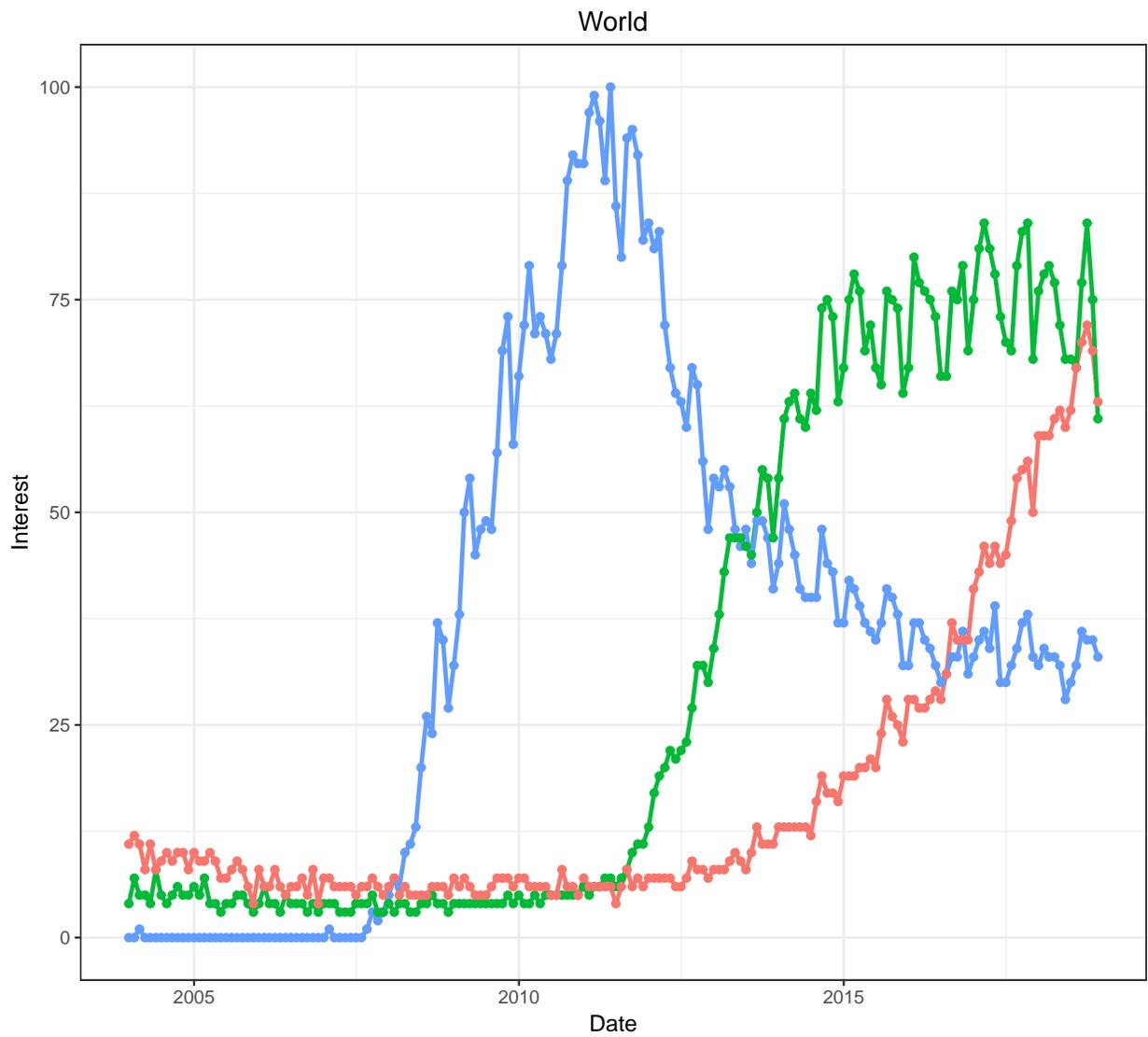}
\caption{Google trends for the terms \lq\lq Data Science'' (red), \lq\lq Big Data'' (green), and \lq\lq Cloud Computing'' (blue) for global queries. The scores assigned by Google Trends on the \lq\lq interest'' ordinate express the popularity of that term over a specified time range, based on the absolute search volume for a term, relative to the number of searches received by Google. The scores have no direct quantitative meaning. For example, two different terms that have been searched 1000 and 20000 times, respectively, could achieve a score of 100. This is because the scores have been scaled between 0 and 100, and a score of 100 always represents the highest relative search volume. Yearly scores are calculated on the basis of the average relative daily search volume within the year.}
\label{fig:gtrendworld}
\end{figure}

\begin{figure}[!ht]
\centering
\includegraphics[width = \textwidth]{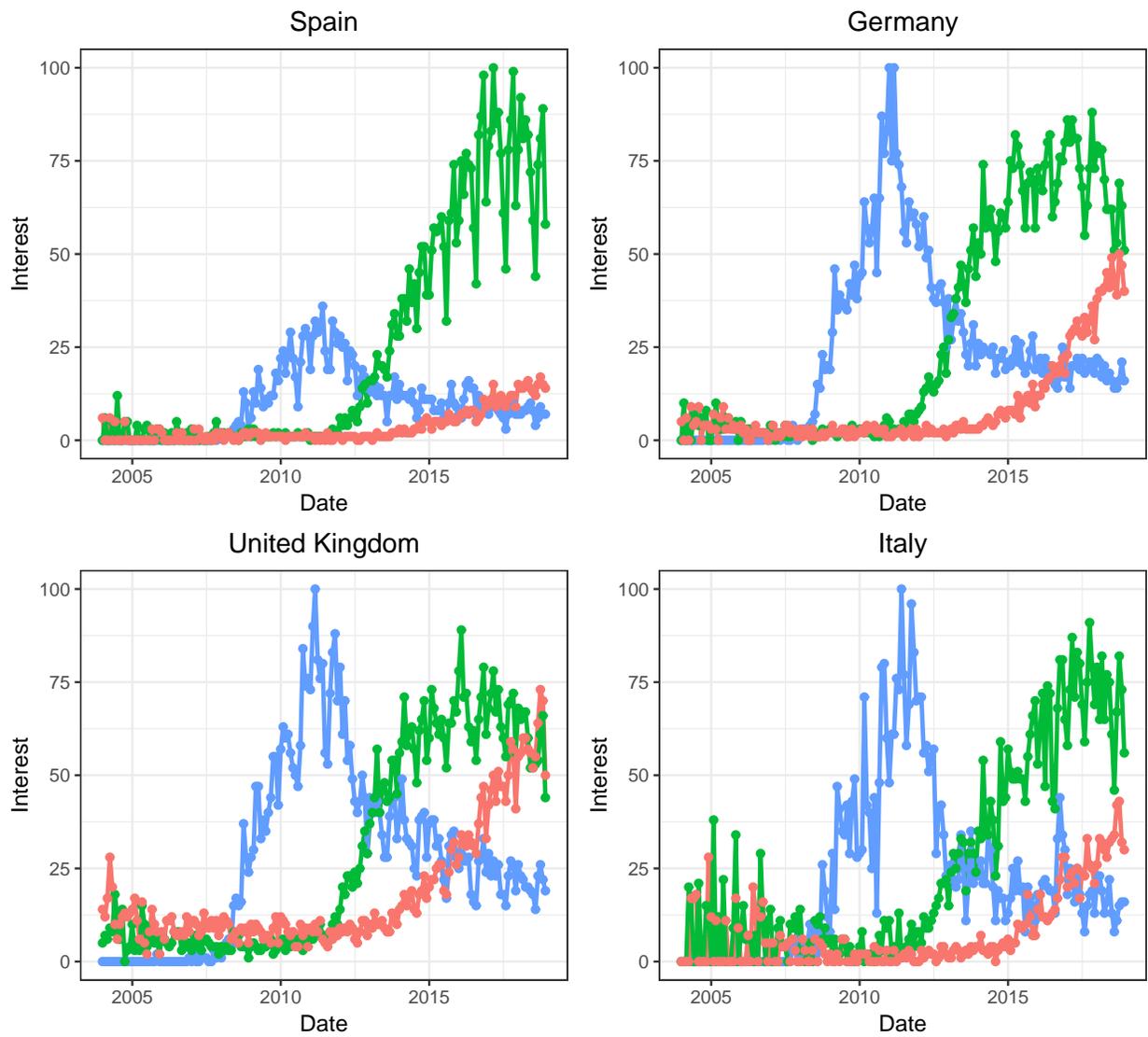}
\caption{Google trends for the terms \lq\lq Data Science'' (red), \lq\lq Big Data'' (green), and \lq\lq Cloud Computing'' (blue) for some countries of Europe.}
\label{fig:gtrendeurope}
\end{figure}

\begin{figure}[!ht]
\centering
\includegraphics[width = \textwidth]{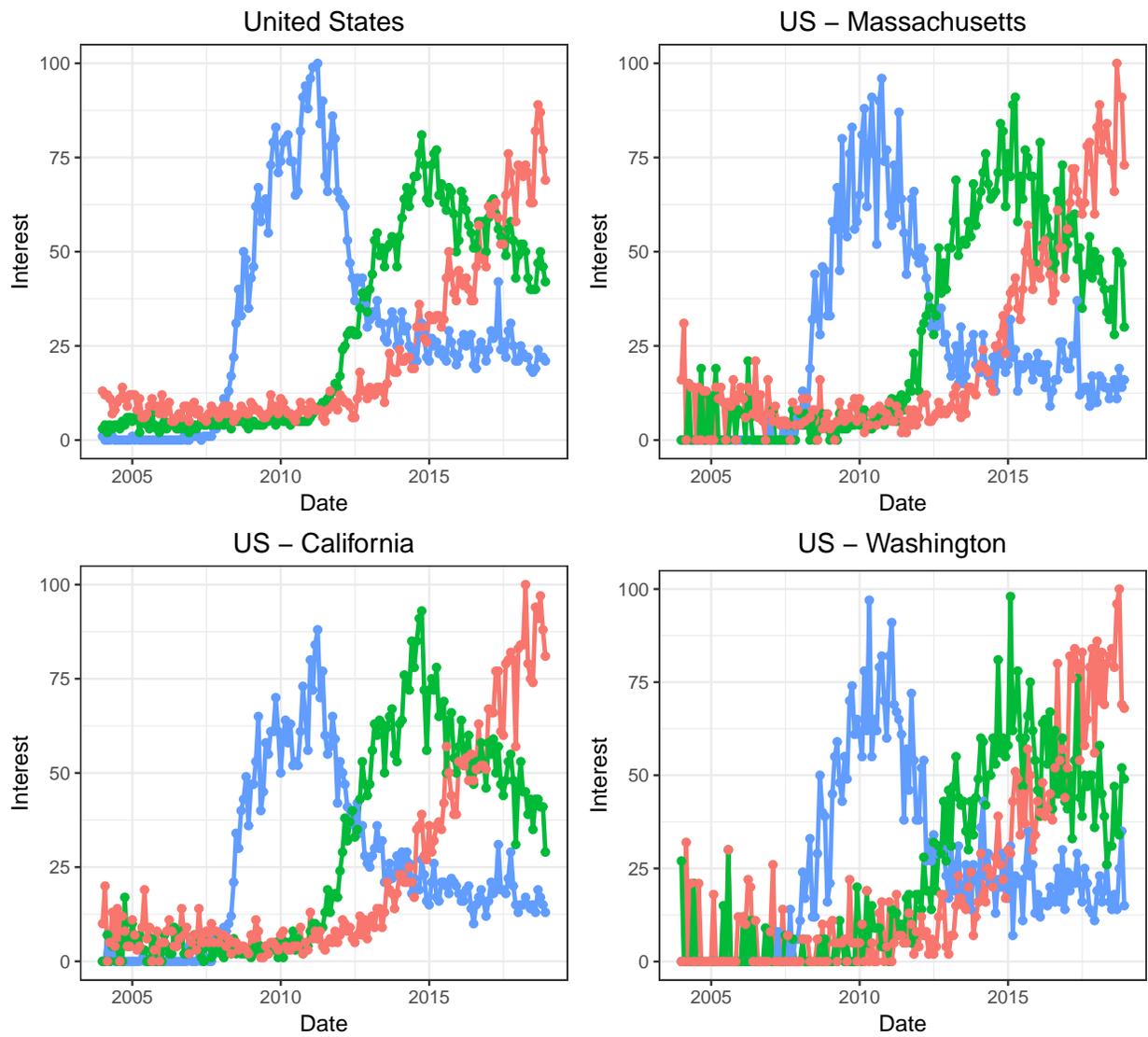}
\caption{Google trends for the terms \lq\lq Data Science'' (red), \lq\lq Big Data'' (green), and \lq\lq Cloud Computing'' (blue) for United States and some of its states.}
\label{fig:gtrendusa}
\end{figure}

\begin{figure}[!ht]
\centering
\includegraphics[width = \textwidth]{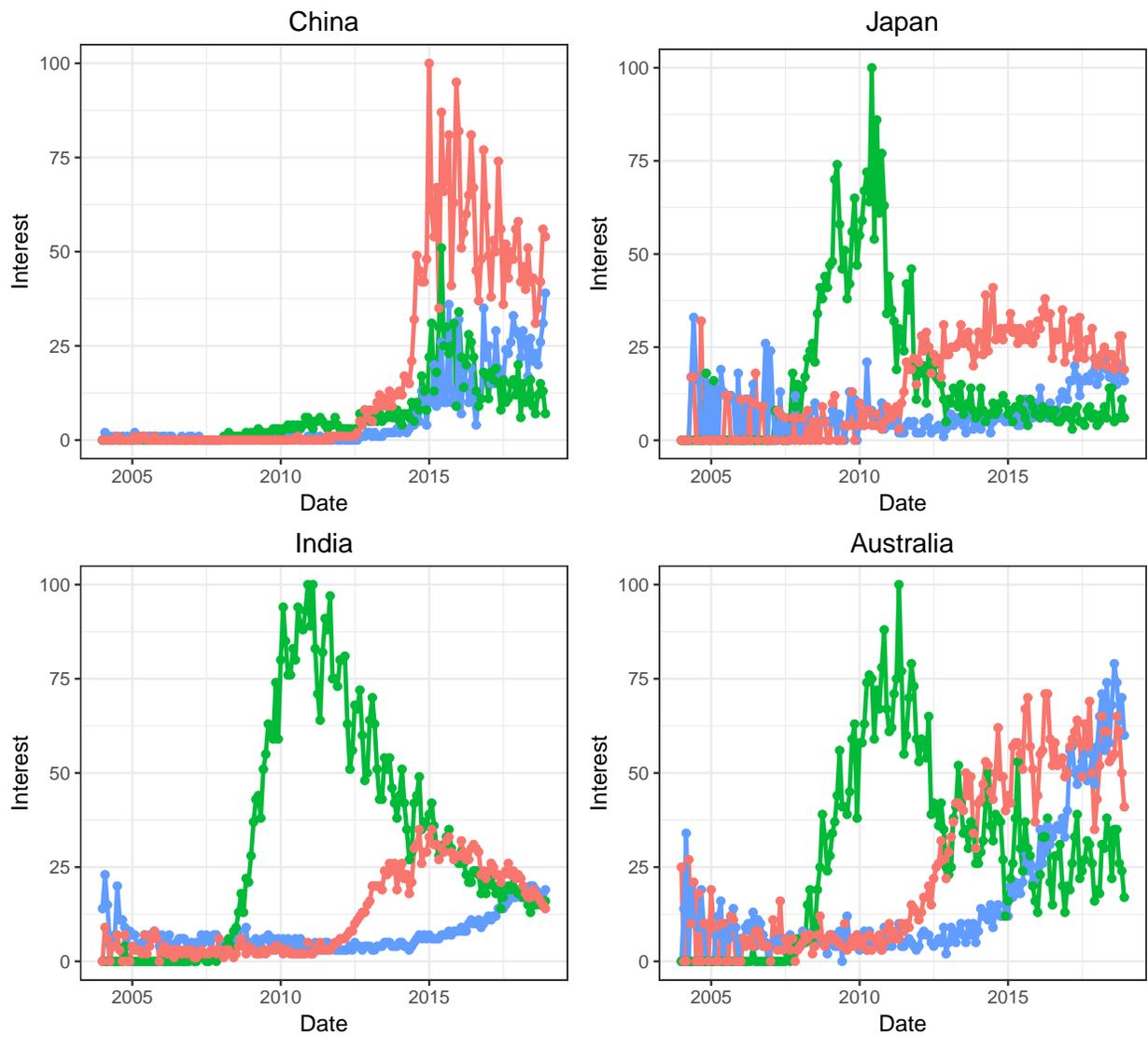}
\caption{Google trends for the terms \lq\lq Data Science'' (red), \lq\lq Big Data'' (green), and \lq\lq Cloud Computing'' (blue) in some countries of Asia and in Australia.}
\label{fig:gtrendasiaau}
\end{figure}

To visualize the progress of the terms \lq\lq Data Science'',  \lq\lq Cloud Computing'', and \lq\lq Big Data'', we present the results obtained both worldwide and in some countries in Europe, the United States (and some of its states), in Asia, and in Australia over time. The results are summarized in Figures \ref{fig:gtrendworld} - \ref{fig:gtrendasiaau}. All the searches were performed using the \texttt{R} package \texttt{gtrendsR} \citep{gtrendsr}, which is an interface for retrieving and displaying the information returned online by Google Trends.

An up-tick in Data Science is not produced until approximately the year 2012. It is precisely in this year that the interest for the term \lq\lq Big Data'' starts to grow at high rate. On the other hand, by the end of 2014 and the beginning of 2015, the trend for searches on \lq\lq Big Data'' begins to stagnate, and we can observe an almost exponentially increasing interest for the term \lq\lq Data Science''. On the other hand, the term \lq\lq Cloud Computing'', had its main boom around 2011, and since then, its influence has been decreasing. 

However, in some countries such as Spain, no real peak for the term \lq\lq Data Science'' is observed until the year 2015. Even though there is also an increase in searches about this concept, the growth is much less pronounced than in other European countries such as Germany, where the interest for \lq\lq Data Science'' is equal to that of \lq\lq Big Data'', or the United Kingdom, where the trend for \lq\lq Data Science'' begins to unseat that of \lq\lq Big Data'' (Figure \ref{fig:gtrendeurope}). The trend is even more pronounced in United States, in particular in some of its states such as Massachusetts or California, where the main universities and research centers are. In these US states, the trend for \lq\lq Big Data'' is decreasing sharply coinciding with a growing interest in \lq\lq Data Science'' (see Figure \ref{fig:gtrendusa}). In other countries such as China, India, or Japan, the pattern of interest on these terms is similar but with a certain slowness with respect to other countries. It seems that the interest in \lq\lq Data Science'' in these countries as well as in Spain, has not yet reached the same level as in other parts of the world (Figure \ref{fig:gtrendasiaau}).

With this search, we reassert the findings presented in \citet{kane2014cleveland}: i) The trend for the term \lq\lq Data Science'' is eclipsing the popularity of the infrastructure on which it is based (cloud computing, big data, computational skills, etc.) specially in the more technological countries; ii) The interest for Data Science is increasing worldwide and it appears that the trend is that this growth will continue in the coming years.	

\subsection*{Journals of Data Science}

\begin{table}[]
\small
\centering
\caption{Current journals in the Data Science field up to 2018.}
\label{table:journals}
\begin{tabular}{@{}lcccc@{}}
\toprule
\multicolumn{1}{l}{\textbf{Journal and website}}                                                                                           & \textbf{Publisher}                                         & \textbf{Scopus} & \textbf{\begin{tabular}[c]{@{}c@{}}Open \\ access\end{tabular}}
& \textbf{\begin{tabular}[c]{@{}c@{}}Bio/health \\ research\\ (explicity)\end{tabular}} \\
\midrule
\begin{tabular}[c]{@{}l@{}}Journal of Data Science\\ \url{http://www.jds-online.com}\end{tabular}                                                & -                                                          & No              & No                                                              & No                                                                                    \\
\midrule
\begin{tabular}[c]{@{}l@{}}International Journal of Data Science and Analytics\\
\url{https://link.springer.com/journal/41060}\end{tabular}      & Springer                                                   & No              & Hybrid                                                          & Yes                                                                                   \\ \midrule
\begin{tabular}[c]{@{}l@{}}Data Science Journal\\ \url{https://datascience.codata.org}\end{tabular}                                              & -                                                          & Yes             & Yes                                                             & Yes                                                                                   \\
\midrule
\begin{tabular}[c]{@{}l@{}}Data Science- Methods, Infrastructure, and Applications\\
\url{https://datasciencehub.net}\end{tabular}               & IOS Press                                                  & No              & Yes                                                             & No                                                                                    \\
\midrule
\begin{tabular}[c]{@{}l@{}}EPJ Data Science\\ \url{https://epjdatascience.springeropen.com}\end{tabular}                                         & Springer                                                   & Yes             & Yes                                                             & Yes                                                                                   \\
\midrule
\begin{tabular}[c]{@{}l@{}}International Journal of Data Science\\
\url{http://www.inderscience.com/jhome.php?jcode=ijds}\end{tabular}           & Inderscience                                               & No              & Hybrid                                                          & No                                                                                    \\ \midrule
\begin{tabular}[c]{@{}l@{}}Advances in Data Science and Adaptive Analysis\\
\url{https://www.worldscientific.com/worldscinet/adsaa}\end{tabular} & \begin{tabular}[c]{@{}c@{}}World\\ Scientific\end{tabular} & No              & Hybrid                                                          & No                                                                                    \\
\bottomrule
\end{tabular}
\end{table}

In this new field, there are only seven scientific journals directly related with the data science (up to 2018); see Table\ref{table:journals}. Notice that we do not consider journals that are only related to Big Data Analysis or Machine Learning for the reasons exposed in Section \ref{sec:intro}.

The goal of the \textit{Journal of Data Science} is to enable scientists to do their research on applied science and through the effective use of data. Regarding the \emph{International Journal of Data Science and Analytics}, the main topics addressed are data mining and knowledge discovery, database management, artificial intelligence (in\-clu\-ding robotics), computational biology/bioinformatics, and business information systems. The related industry sectors are: electronics, telecommunications and IT \& Software. The \textit{International Journal of Data Science and Analytics} brings together researchers, industry practitioners, and potential users of big data, to promote collaborations, exchange ideas and practices, discuss new opportunities, and investigate analytics frameworks. The jour­nal welcomes experimental and theoretical findings on data science and advanced analytics along with their applications to real-life situations. The scope of the \textit{Data Science Journal} includes descriptions of data systems, their publication on the internet, applications and legal issues. All the sciences are covered, including the Physical Sciences, Engineering, the Geosciences, and the Biosciences, along with Agriculture and the Medical Science. The ultimate goal of \textit{Data Science - Methods, Infrastructure and Applications} is to unleash the power of scientific data to deepen our understanding of physical, biological, and digital systems, gain insight into human social and economic behaviour, and design new solutions for the future. Additionally, the \textit{EPJ Data Science} covers a broad range of research areas and applications and particularly encourages contributions from techno-socio-economic systems. Topics include, but are not limited to, human behaviour, social interaction (including animal societies), economic and financial systems, management and business networks, socio-technical infrastructure, health and environmental systems, the science of science, as well as general risk and crisis scenario forecasting up to and including policy advice. Finally, the \textit{International Journal of Data Science} aims to provide a professional forum for examining the processes and results associated with obtaining data, as well as munging, scrubbing, exploring, modelling, interpreting, communicating and visualizing data. Data science takes data in cyberspace as a research object. The goal is an integrated and interconnected process designed to form a common ground from which a knowledge-based system can be built, shared, and supported by professionals from different disciplines. Finally, \textit{Advances in Data Science and Adaptive Analysis} is an interdisciplinary journal dedicated to report original research results on data analysis methodology developments and their applications, with a special emphasis on the adaptive approaches. The mission of the journal is to elevate data analysis from the routine data processing by traditional tools to a new scientific level, which encourages innovative methods development for data science and its scientific research and engineering applications.

As we can see in Table \ref{table:journals}, not all the journals listed above explicitly include health data science and none of them is exclusively dedicated to this area. Following, we provide a proper description of what we consider health or biomedical data science.
\section{Data Science in the biomedical field}	
\label{sec:biomedical}

A Biomedical Data Scientist should be quantitatively trained including a comprehensive and rigorous proficiency of statistical principles and those computing skills to handle massive and complex data. He/she has to be able to manage and analyze health data to solve emerging problems in public health and biomedical sciences and to learn how to interpret their findings.

Health data refers to data that come from the biomedical sciences, public health, and any other area related to the \lq\lq bio'' sciences. Examples are data sets from clinical trials, observational studies, genomics and other omics studies, medical records, health care programs, or environmental programs.

Health-related data are also a good example of the legal and ethical concerns that should be taken into consideration regarding sensitive personal data (medical records, genomic profiles, etc.) or digital epidemiology in the context of public health. Thus, ensuring compliance with ethical policies, adequate informed consents, and data use agreements are essential when sharing information and collaboratively using data \citep{gomez2016big}.

\begin{figure}[!h]
\centering
\includegraphics[scale = 0.5]{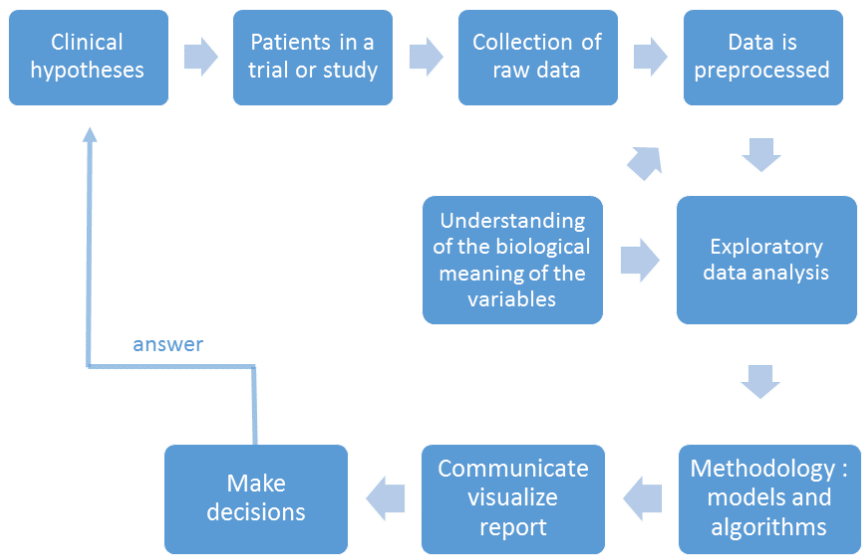}
\caption{Healthcare field process in which a data scientist is involved.}
\label{fig:datascience}
\end{figure}

Several researchers have shown that the use of adequate tools for analyzing these data could help to reduce mortality rates, health care costs, or even improve the living conditions of patients \citep{reddy2015healthcare}. This shows up the importance of data science in the biomedical area, also called Biomedical Data Science (BDS) and is the main aim of this manuscript. It can be defined as the combination of statistical knowledge, computational skills, and expertize in the health science with the aim of answering important questions or hypotheses, as we show in Figure \ref{fig:datascience}. This figure, based on a similar diagram in \lq Doing Data Science' by \citet{kanaan2014doing}, shows the process in which a data scientist needs to be involved and exhibits why data science, as a field, needs to join forces and cooperate with different areas.

According to the field of study and previous clinical hypotheses, patients who meet the inclusion criteria are recruited into a trial or study, and the raw data such as clinical parameters, demographics, or omics data is collected.  Following, preprocessing of the data is done with the objective of cleaning and preparing the data set for exploratory analysis extracting important descriptive statistics. The next step is to find the right methodology to provide an answer to the specific questions for this problem. First, existing models have to be explored for their adequacy to the data and the relevant question. Then, the chosen method has to be implemented and developed. In many occasions new methods have to be developed or old methods have to be adequately adapted. Moreover, the data scientist needs to be able to communicate properly and clearly the results obtained by means of reports and graphical tools. For these reasons, statistics is a fundamental part of the decision making process, it helps draw conclusions and answer the clinical hypotheses. As an added value, the data scientist understands the biological problem, know the biological meaning of the main variables and has to manage a common vocabulary with the rest of the team.

Due to the very likely possibility of potential statistical pitfalls when adapting or developing the chosen methodology, the data scientist should be reliable, coherent and a guide to follow. A small list of these pitfalls are biased samples, overgeneralization, spurious correlation, prediction performance, incorrect analysis choices, and violation of the assumptions for an analysis. A good example of biased samples is cited by \citet{crawford2013hidden} and shows the data collected in the city of Boston through the StreetBump smartphone app, created with the objective of solving the problem with potholes in this city. This app passively detects bumps by recording the accelerometers of the phone and GPS data while driving, instantly reporting them to the traffic department of the city. Thus, the city could plan their repair and the management of resources in the most efficient possible way. However, one of the problems observed was that some segments of the population, such as people in lower income groups, have a low rate of smartphone use, a rate that is even lower in the older residents, where smartphone penetration is as low as 16\%. Therefore, these data provide a big but very biased sample of the population of potholes in the city, with the consequent impact on the understimation of the number of potholes in certain neighborhoods and the deficient management of resources. Thus provide a clear instance when having large amounts of data is not synonymous of quality and using the data to solve a problem might result in unfair and not cost effective policies.

Statistical thinking is the central element to avoid the above-mentioned pitfalls. It requires a non-trivial understanding of the real-world problem and the population for whom the research question is relevant. It involves judgements such as those about the relevance and representativeness of the data, about whether the underlying model assumptions are valid for the data at hand, and about causality and the role of confounding variables as possible alternative explanations for observed results. In fact, an essential component of good statistical thinking is the ability to interpret and communicate the results of a statistical analysis so non statisticians can understand the findings \citep{greenhouse2013statistical}. In \textit{The Seven Pillars of Statistical Wisdom} \citet{stigler2016seven} summarizes Statistical reasoning as an integral part of modern scientific practice and sets forth the foundation of statistics around seven principles. Stigler's second pillar, Information,  challenges the importance of \lq\lq big data'' by noting that observations are not all equally important: the amount of information in a data set is often proportional to only the square root of the number of observations, not the absolute number.

\subsection{Biomedical Data Science in the Web of Science}

Similar to the search presented in Section \ref{sec:dsglobalimpact}, we have analyzed the number of publications associated with \lq\lq Data Science'', \lq\lq Big Data'', and \lq\lq Cloud Computing'' in several countries and along the last fourteen years, using Web of Science (\url{https://clarivate.com/products/web-of-science/}). The countries considered were Australia, China, Germany, India, Italy, Japan, Spain, the United Kingdom, and the United States. Notice that \lq\lq publication'' refers to articles, reviews, clinical trials, case reports, and books. Moreover, only topics related with the biomedical area, such as Oncology, Respiratory System, or Pediatrics, were considered. The publication counts were obtained at the beginning of 2019 and are presented in Table \ref{tab:awos} (Appendix \ref{sec:appendixA}).

From the publication counts presented in Table \ref{tab:awos} (see Appendix \ref{sec:appendixA}), we can conclude that the number of biomedical publications has increased during the last years in the countries considered. Moreover, as might be expected, the number of publications associated with \lq\lq Data Science'' is much larger than the number of publications associated with the topics \lq\lq Big Data'' and \lq\lq Cloud Computing''. In fact, the publications associated with Data Science represent more than $95\%$ of all the publications analyzed, regardless of the country considered. Most noteworthy is the tremendous increase of record counts in China: 917 publications were registered in 2004, and this number has increased to 12013 in 2017; that is, an increase of more than 10000 publications in only 13 years. Furthermore, the case of Spain is also remarkable because the presence of publications associated with \lq\lq Data Science'' is much lower than in other European countries like Germany, Italy, or the United Kingdom. For instance, in 2017 the number of publications in the United Kingdom and in Germany is approximately three times and twice as high as in Spain, respectively. Although the comparison is not immediate because the population of United Kingdom and Germany is more two time as high as in Spain. 

On the other hand, the presence of publications about \lq\lq Cloud Computing'' in the biomedical area is really low: until after 2010, very low number of publications were registered in any of the countries considered. Even in 2017 the number of publications was low compared with the other topics. We can, hence, state that the use of Cloud Computing techniques is not widespread among researchers in the field of Biomedicine. Finally, the explosion of \lq\lq Big Data'' in the last years, seems to have an effect in the Biomedical research because the number of publications in this topic has increased each year in the countries considered.  For example, in Australia the number of publications about \lq\lq Big Data'' in 2007 was 17 as compared to 131 publications in 2017, that is, an increase greater than $670\%$. It is clear that Big Data techniques have been very useful in order to solve biomedical problems.

\subsection{Multidisciplinary environment for Biomedical Data Science}

The confluence of science, technology, and medicine in our dynamic digital era has spawned new data applications to develop prescriptive analytics, to improve healthcare personalization and precision medicine, and to automate the reporting of health data for clinical decisions \citep{bhavnani2016data}. As we mentioned before, several biomedical research institutes are involved in the data science process working on complex data bases in the areas of genomic and proteomic data analysis, infectious and immunological diseases, new therapies in cancer, hormones and cancer, genetics, cellular biology, among others. Most of the research studies need data science techniques to deal with these data sets. Those data science studies that are usually characterized by complex structures or large numbers of variables, require a multidisciplinary environment with biomedical informatics, bioinformatics, biostatisticians, and clinicians. This environment brings together statistics, computer sciences, and computational engineering, and aims to provide a methodologically correct analysis.

Biomedical Data Science can be applied in many different areas such as personalized medicine, genomic research, gene expression analysis, or in cancer drug studies, among others. Following, we present some examples of applications.

Personalized medicine is a medical approach in which patients are stratified in subgroups according to their individual characteristics (genomic alterations, lifestyles, diagnostic markers, clinical profile, response to treatments). With abundant and detailed patient data, medical decisions, such as diagnostic tests or treatments, may be personalized and addressed to these subgroups of patients and not to the whole population. The advantages of personalized medicine are evident: more effective use of therapies and reduction of adverse effects, early disease diagnosis and prevention by using biomarkers, among others. A well-known example is the treatment with trastuzumab (Herceptin, a breast cancer drug) that can only be administered if the HER2/neu receptor is overexpressed in tumor tissue because the drug interferes with this receptor. Another example of those personalized predictions can be the survival probabilities predicted for a future level of a longitudinal biomarker recorded. The joint model approaches to study the association between a longitudinal biomarker and survival data provides dynamic predictions for survival probability coming from the effect of the longitudinal biomarker taken until time $t$, which can be updated when the patient has new information \citep{rizopoulos2011dynamic, rizopoulos2012joint}.

Data science helps to examine health disparities. Research examining racial and ethnic disparities in care among older adults is essential for providing better quality care and improving patient outcomes. Yet, in the current climate of limited research funding, data science provides the opportunity for gerontological nurse researchers to address these important health care issues among racially and ethnically diverse groups, groups typically under-represented and difficult to access in research \citep{chase2016examining}.

Other example is to use data science for clinical decision making. Clinical laboratories contribute towards the screening, diagnosis and monitoring of many types of health conditions. While it is believed that diagnostic testing may account for just 2\% - 4\% of all healthcare spending, it may influence 60\% - 80\% of medical decision-making. The work of \citet{espasandin2018geographical} is an example of BDS where a very recent extension of the distribution regression model introduced by \citet{klein2015bayesian} is applied to a data set of blood potassium concentrations from patients across a Spanish region. The main aim of this manuscript was to determine if geographical differences possibly attributed to pre analytical factors could be detected.

The development of automated workflows that can capture and memorialize extensive experimental protocols, aiding in reproducibility as well as taking data analysis to a new level \citep{ludascher2006scientific} is a central data science technique. Workflows help support and accelerate scientific discoveries in biomedical research by eliminating the burden of dealing with time-consuming data and software integration. This approach fundamentally frees researchers to concentrate on the scientific questions at hand instead of addressing technical issues involved in setting up, executing, and validating the computational pipeline \citep{amaro2016drug}.
	
We can found applications in many other fields. For instance, while studying the consequences of the analytical treatment interruption in HIV-infected patients, \citet{alarcon2018multiple} present a method to fit a mixed effects Cox model with interval-censored data to study the viral rebound of HIV. The proposal is based on a multiple imputation approach that uses the truncated Weibull. The authors addressed the fact of having data from eight different studies based on different grounds.
	
Another application is to quantify spatio-temporal effects to graft failures in organ transplantation. The transplantation of solid organs is one of the most important accomplishments of modern medicine. Yet, organ shortage is a major public health issue. Using data science, the research can investigate early graft failure time. When an organ becomes available from a deceased donor, the allocation policies such as medical urgency, expected benefit and geographical constraints (distance between donor and recipient) are applied to people in the waiting list to select a match. Allocation policies regard the survivability of the organ outside the human body, namely, the cold ischemic time, as an important factor since it is associated with the quality degradation of the organ. Besides, the distance is an important factor on these decisions given that the farther the distance from the donor hospital to the transplant center, the worse might be the quality of the organ \citep{pinheiro2016data}.
	
We can even relate data science with mental health. Mental disorders are arguably the greatest \lq hidden' burden of ill health, with substantial long-term impacts on individuals, carers and society. People with these conditions are often socially excluded and less likely to participate in research studies or remain in follow-up. Complexities around defining diagnoses present particular challenges for mental health research. Richly annotated, longitudinal data sets matched to data science analytics offer an unprecedented opportunity for more robust diagnostics, and also the prediction of outcome, treatment response, and patient preferences to inform interventions \citep{mcintosh2016data}.

Many more examples of BDS are expected to arise in any other field related to health or bio sciences in the near future.

\subsection{Standardization of information}

From the above, we could say that one of the main objectives of Data Science in Biomedicine is to generate valid knowledge through better structuring in the procedures for extracting, analyzing and processing data obtained in health and environmental research, supporting the transfer of their results to society. All these disciplines share common goals in terms of improving the quality of life of the people through actions in the promotion of health and in the prevention of disease.

A major challenge that exists in the healthcare domain is the \lq\lq data privacy gap" between medical researchers and computer scientists. Medical researchers have natural access to healthcare data because their research is paired with a medical practice. Acquiring data is not quite as simple for computer scientists without a proper collaboration with a medical practitioner. There are barriers in the acquisition of data. Many of these challenges can be avoided if accepted protocols, technologies, and safeguards are in place.

On the other hand, people to whom the research efforts are addressed and those responsible for funding agencies need to ensure that research output are used to maximize knowledge and potential benefits.
Sharing the data ensures that these are available to the research community, which accelerates the pace of discovery and enhances the efficiency of the research. Believing on these benefits, many initiatives actively encourage investigators to make their data available.

Widely available crowd-sourcing programs such as PatientsLikeMe (\url{www.patientslikeme.com}) have amassed participation from more than 400 thousand patients across 2500 disease conditions who actively share health related data on an open and online platform that tracks and collects important patient-reported outcomes. The United Kingdoms BioBank is a large-scale biomedical data set containing detailed phenotypic, genotypic, and multimodal imaging findings to determine the genetic and nongenetic determinants of health and disease in a contemporary cohort of more than 500000 participants. Available through open access, research collaborations have advanced our knowledge in the risk prediction of cardiovascular, psychiatric, and cerebrovascular diseases and have identified important anthropometric and genetic traits of metabolic health including diabetes mellitus and obesity.

The objectives for these kind of initiatives are similar to the established data sources such as census and public health data sets, or standardized patient registries such as the National Cardiovascular Data Registry, where data are structured and aggregated. The objective is to monitor population trends, develop guideline-based care, and infer changes to healthcare policy, new citizen science and crowd-sourcing initiatives aim to leverage public and patient participation to collect health data and vital statistics through new massive open, and online data repositories \citep{bhavnani2016data}.

Since 2003, the National Institutes of Health (NIH)  has required a data sharing plan for all large funding grants. Similarly, some journals are also requiring the deposit of data and other research documentation associated with published articles \citep{borgman2012conundrum, piwowar2007sharing}.

In May, 2010, the Wellcome Trust and the Hewlett Foundation convened a workshop in Washington, DC, to explore how funders could increase the availability of data generated by their funded research, and to promote the efficient use of those data to accelerate improvements in public health \citep{walport2011sharing}. In this meeting, funders agree to promote greater access to and use of data in ways that are: equitable, ethical and efficient. Equitable refers to recognizing those researchers who generate the data, other analysts reusing these data,  meanwhile population and communities expect health benefits arising from research. It should protect the privacy of individuals. Healthcare data is obviously very sensitive because it can reveal compromising information about individuals. Several laws in various countries explicitly forbid the release of medical information about individuals for any purpose, unless safeguards are used to preserve privacy. Finally, it should improve the quality and value of research, and increase its contribution to improving public health.

In June 2018, the NIH releases its first Strategic Plan for Data Science (\url{https://www.nih.gov/news-events/news-releases/nih-releases-strategic-plan-data-science}). In this plan, \lq\lq NIH addresses storing data efficiently and securely; making data usable to as many people as possible; developing a research workforce poised to capitalize on advances in data science and information technology; and setting policies for productive, efficient, secure, and ethical data use. This plan commits to ensuring that all data-science activities and products supported by the agency adhere to the FAIR principles, meaning that data be Findable, Accessible, Interoperable, and Reusable" \citep{wilkinson2016fair}.

\section{Conclusions}
\label{sec:discus}
Motivated by the remarkable increase of the number of publications on Data Science in the past few years, the purpose of this work has been to study the impact of Data Science in the area of biomedicine. 

With this objective in mind, we have carried out a search of the terms \lq\lq Data Science" along with \lq\lq Big Data'' and \lq\lq Cloud Computing" using Google Trends until November 2018. While Big Data represents the information assets characterized by a high volume, velocity and variety to require specific technology and analytical methods for its transformation into value \citep{de2015big}, Cloud Computing enables ubiquitous, convenient, on-demand network access to a shared pool of configurable computing resources (e.g., networks, servers, storage, applications, and services) that can be rapidly provisioned and released with minimal management effort or service provider interaction \citep{mell2009nist}. Big Data and Cloud Computing were chosen since they are somewhat related to computing movements and they help to put the \lq\lq Data Science'' search-traffic  into  perspective \citep{kane2014cleveland}. According to our search results, in the last years more and more publications in the area of biomedicine make use of the term \lq\lq Data Science'', however, there are large differences among the countries considered.

We have also listed the main journals only related to Data Science to point out the increasing importance of Data Science. However, not all of the journals presented explicitly include Biomedical Data Science (BDS) as their main areas of research. In addition, we have stepped ahead of the contemporary definition of Data Science, directly related to the economics or business world, describing the Data Science in the Biomedical field. We understand BDS as the interdisciplinary field that encompasses the study and pursuit of the effective use of biomedical data, information, and knowledge for scientific inquiry, problem-solving, and decision-making, driven by efforts to improve human health. It investigates and supports reasoning, modelling, simulation, experimentation, and translation across the spectrum, from molecules to individuals to populations. 

We strongly believe that the importance of Biomedical Data Science will continue increasing in the near future due to nowadays' possibilities to record enormous quantities of data and the technical facilities to process them. Statistical thinking and knowledge will play a key role in the correct analysis of such data.

\section*{Acknowledgements}

This research was partially supported by the projects: MTM2015-64465-C2-1-R,  MTM2014-52975-C2-1-R, MTM2016-76969-P cofinanced by the Ministry of Economy and Competitiveness (SPAIN), and the projects MTM2017-83513-R and MTM2017-90568-REDT cofinanced by the Ministry of Economy and Competitiveness (Spain), all them cofinanced by the European Regional Development Fund (FEDER). This work was also supported by grants from the Galician Government (ED341D-R2016/032 and ED431C 2016-025),  and by grants from the Carlos III Health Institute, Spain (PI16/01395; PI16/01404; RD16/0007/0006 and RD16/0017/0018), and 2017 SGR 622 (GRBIO) from the Departament d’Economia i Coneixement de la Generalitat de Catalunya (Spain). Work of M. Conde-Amboage has been supported by post-doctoral grant from Ministry of Culture, Education and University Planning and Ministry of Economy, Employment and Industry of Galician Government. The authors want to thank the network BIOSTATNET for many fruitful discussions. Additionally, Yovaninna Alarc\'on-Soto wants to thank to CONICYT for her scholarship.

\vspace*{1pc}

\appendix

\section{Publications on Data Science, Cloud Computing and Big Data in the last fourteen years}
\label{sec:appendixA}

\begin{table}[]
\centering
\caption{Number of publications associated with the topics \lq\lq Data Science'' (denoted by DS), \lq\lq Big Data'' (denoted by BD) and \lq\lq Cloud Computing'' (denoted by CC) in different countries from 2004 to 2018.}
\label{tab:awos}
\begin{minipage}{\textwidth}
\renewcommand{\thefootnote}{\thempfootnote}
\begin{tabular}{ccrrrrrrrrr}
\toprule
Year & Topic & USA & UK & Japan & Germany & Australia & Spain & Italy & India & China \\
\midrule
2004 & DS & 13942 & 3491 & 1551 & 2845 & 1400 &  977 & 1949 & 401 &  917 \\
     & BD &   129 &   28 &   18 &   43 &   10 &   18 &   10 &   2 &   21 \\
     & CC &    28 &    4 &    0 &    4 &    0 &    1 &    4 &   0 &    1 \\
2005 & DS & 16124 & 4272 & 1762 & 3558 & 1556 & 1254 & 2364 & 526 & 1232 \\
     & BD &   162 &   37 &   19 &   45 &   16 &   21 &   10 &   3 &   29 \\
     & CC &    31 &    5 &    2 &    3 &    0 &    0 &    3 &   1 &    2 \\   
2006 & DS & 16296 & 4356 & 1671 & 3602 & 1719 & 1313 & 2373 & 553 & 1405 \\ 
     & BD &   116 &   52 &   13 &   35 &   16 &   26 &   21 &   4 &   48 \\ 
     & CC &    24 &    2 &    3 &    9 &    0 &    1 &    2 &   2 &    2 \\ 
2007 & DS & 16208 & 4421 & 1702 & 3441 & 1717 & 1305 & 2567 & 592 & 1656 \\
     & BD &   162 &   43 &   25 &   46 &   17 &   22 &   20 &   9 &   60 \\
     & CC &    27 &    3 &    3 &    2 &    0 &    3 &    4 &   2 &   13 \\
2008 & DS & 17799 & 4798 & 1792 & 3769 & 1923 & 1542 & 2590 & 715 & 2085 \\
     & BD &   190 &   48 &   20 &   57 &   28 &   29 &   21 &   5 &   62 \\
     & CC &    35 &    5 &    2 &    9 &    1 &    4 &    1 &   3 &   12 \\
2009 & DS & 19181 & 5329 & 2051 & 4325 & 2264 & 1879 & 3091 & 851 & 2694 \\
     & BD &   166 &   59 &   18 &   50 &   30 &   29 &   28 &  11 &   75 \\
     & CC &    37 &    9 &    3 &   15 &    5 &    4 &    3 &   1 &   11 \\
2010 & DS & 21162 & 5920 & 2159 & 4763 & 2690 & 2133 & 3309 &1051 & 3437 \\
     & BD &   206 &   66 &   25 &   74 &   28 &   18 &   31 &  14 &  103 \\
     & CC &    70 &    9 &    9 &   12 &    6 &    8 &    4 &   7 &   17 \\
2011 & DS & 23547 & 6592 & 2404 & 5205 & 2985 & 2373 & 3668 &1170 & 4458 \\
     & BD &   242 &   58 &   23 &   65 &   43 &   33 &   38 &  11 &  131 \\
     & CC &   114 &    7 &   11 &   27 &    6 &    4 &    9 &   7 &   57 \\
2012 & DS & 25815 & 7726 & 2801 & 5778 & 3461 & 2744 & 4037 &1394 & 6008 \\
     & BD &   242 &   66 &   12 &   79 &   40 &   37 &   44 &  16 &  112 \\
     & CC &   133 &   16 &   13 &   22 &   13 &   15 &    9 &  14 &   49 \\
2013 & DS & 27258 & 8020 & 2822 & 6027 & 3940 & 2989 & 4527 &1593 & 7159 \\
     & BD &   353 &   97 &   36 &  100 &   63 &   33 &   40 &  22 &  171 \\
     & CC &   146 &   37 &   17 &   44 &   18 &   25 &   13 &   7 &   82 \\
2014 & DS & 27894 & 7891 & 2792 & 6195 & 4101 & 3156 & 4686 &1747 & 8796 \\
     & BD &   508 &  127 &   44 &  115 &   80 &   55 &   60 &  24 &  214 \\
     & CC &   171 &   36 &   12 &   27 &   26 &   23 &   27 &  17 &  118 \\
2015 & DS & 29141 & 8750 & 2943 & 6356 & 4525 & 3208 & 4883 &1870 &10485 \\ 
     & BD &   691 &  186 &   48 &  148 &  103 &   72 &   67 &  44 &  302 \\ 
     & CC &   171 &   41 &   16 &   44 &   25 &   38 &   33 &  31 &  112 \\ 
2016 & DS & 28825 & 8703 & 2905 & 6598 & 4788 & 3203 & 4886 &2038 &10789 \\
     & BD &   805 &  230 &   48 &  191 &  123 &   79 &  102 &  71 &  319 \\
     & CC &   212 &   53 &   21 &   48 &   33 &   38 &   34 &  66 &  140 \\
2017 & DS & 29051 & 8622 & 2858 & 6480 & 4617 & 3312 & 4919 &1962 &11727 \\
     & BD &   879 &  266 &   63 &  201 &  131 &   95 &   91 &  69 &  482 \\
     & CC &   221 &   47 &   16 &   44 &   45 &   55 &   50 &  48 &  209 \\
2018 & DS & 27397 & 8361 & 3014 & 6299 & 4678 & 3206 & 4927 &1942 &12013 \\
     & BD &   917 &  260 &   58 &  202 &  143 &  112 &  129 &  89 &  540 \\
     & CC &   230 &   46 &   19 &   41 &   30 &   42 &   33 &  83 &  201 \\
\bottomrule
\end{tabular}
\end{minipage}
\end{table}

\clearpage
\bibliographystyle{apacite}

\bibliography{biblio}

\end{document}